\documentclass[twocolumn,showpacs,showkeys,superscriptaddress,aps]{revtex4}

\usepackage{amsmath}
\usepackage{bbold}
\usepackage{amsfonts} 
\usepackage{amssymb}
\usepackage{pbsi}
\usepackage[T1]{fontenc}
\usepackage{hyperref}
\usepackage{xcolor}
\usepackage{graphicx}
\usepackage{subfig}
\usepackage{float}
\usepackage{soul}
\usepackage{babel}

\usepackage{color}

\begin{document}

\title{Ultrarelativistic outflows in asymmetric magnetic reconnection}

\author{Maricarmen A. Winkler}
\email{maricarmen.winkler@edu.uai.cl}
\affiliation{Facultad de Ingenier\'ia y Ciencias,
Universidad Adolfo Ib\'a\~nez, Santiago 7491169, Chile.}

\author{Felipe A. Asenjo}
\email{felipe.asenjo@uai.cl}\affiliation{Facultad de Ingenier\'ia y Ciencias,
Universidad Adolfo Ib\'a\~nez, Santiago 7491169, Chile.}

\begin{abstract}
We present a  one-fluid  pair plasma magnetohydrodynamical model for  asymmetric relativistic magnetic reconnection that incorporates the thermal-inertial effects of the plasma. We find the general scaling relation for the reconnection rate in a Sweet-Parker-type configuration. However, we show that under a specific highly asymmetric scenario, this magnetic reconnection process can produce ultrarelativistic plasma outflows, with velocities surpassing those of the inflow particles, and also, those
found in symmetric cases. We highlight the significance of the asymmetry in enhancing particle acceleration and energy release.
\end{abstract}


\maketitle

\section{Introduction}

Magnetic fields are ubiquitous across the universe, playing a fundamental role in diverse plasma environments ranging from laboratory experiments to astrophysical systems such as solar wind, flares, magnetospheres, and intergalactic media. The presence of magnetic field lines and their interactions are responsible for many of the processes that we are able to observe and study \cite{cveji2022,grasso2001,ryu2008,Ryu2012,Burlaga2001}. 
Magnetic reconnection is an elemental phenomenon in highly conducting plasmas, where a varying magnetic field can affect the way charged particles move and vice versa. Under certain conditions, magnetic field lines carried along a charged fluid rearrange, altering the configuration of  plasmas \cite{book_priest2000magnetic,book_biskamp2000magnetic,book_gonzalez2016magnetic}. This process converts magnetic energy into heat and kinetic energy, powering several phenomena in space, astrophysical and laboratory plasmas such as solar wind, evolution of magnetospheres in stars, accretion disks and fusion plasmas, to name a few \cite{zweibel2009,mckinney2011,Zhang2011,Sironi2021}.

Magnetic reconnection has been the subject of many theoretical or simulational studies, where slow and fast reconnection is treated. Scaling equations and reconnection rates are analyzed to further understand this process, yet most of these studies consider nonrelativistic plasmas \cite{ni2018,Uzdensky2011,cassak2009,Kadowaki2021,Yang2020}. It has been shown over the years that in magnetically dominated environments, relativistic effects have to be considered since the magnetic energy density surpasses the rest mass energy density, meaning that the Alfvén speed of the wave approximates the speed of light \cite{bessho2012fast,takahashi2011scaling,blackman1994}. Relativistic reconnection can be seen in pulsar winds, jets from active galactic nuclei and pulsar magnetospheres \cite{lyutikov2003explosive,uzdensky2013physical,mckinney2012reconnection}. 

These previous works have also been developed under a resistive  relativistic magnetohydrodynamical model (resistive RMHD), but it has been shown that thermal-inertial effects can be considered to further enhance the reconnection process \cite{asenjocomisso2014}.
Studies have also mostly focus their analysis in a symmetrical scenario, where the magnetic field strength, density, inflow and outflow plasma velocities, and temperatures are all equal for the two reconnecting plasmas. However, it has been frequently observed that different inflow conditions can occur in the heliosphere and near-Earth systems \cite{samadi2017,swisdak2003diamagnetic,malakit2010scaling}. This raises important questions about the role of asymmetric inflow conditions in relativistic reconnection, particularly in extreme astrophysical contexts. 

In this work, we extend previous models \cite{mbarek2022,CassakShay} by incorporating thermal-inertial effects into a general framework of asymmetric relativistic magnetic reconnection. Using a one-fluid RMHD model for pair plasmas, we derive a general scaling relation for the reconnection rate under asymmetric inflow conditions and analyze how these asymmetries modify the dynamics of the process. As a main result, we demonstrate that a highly (very specific) asymmetric configuration can give rise to ultrarelativistic outflows, with velocities significantly exceeding those found in symmetric cases. For this case, the outflow plasma velocities surpass by the far the velocities of the inflow plasma.
This provides a novel mechanism for extreme particle acceleration, with potential applications to high-energy astrophysical systems.


To estimate the magnetic reconnection rate in the general case of asymmetric inflow conditions in a relativistic pair plasma when thermal-inertial effects are considered, we use a one-fluid model based on a relativistic two-fluid approximation  proposed by Koide \cite{koide2009}. In this case, relativistic magnetohydrodynamic equations (RMHD) can be used for a pair plasma with density $\rho$, number density $n$, four-velocity $U_\mu$, satisfying $U_\mu U^\mu =\eta_{\mu\nu} U^\mu U^\nu=-1$, four-current $J^\mu$, and metric signature $\eta_{\mu\nu}=\left(-1,1,1,1\right)$. The plasma continuity equation is written as 
\begin{align}
    \partial_\mu &\left(\rho U^\mu\right)= 0 \label{eq_cont} \ , 
\end{align}
while the generalized momentum equation is
\begin{align}
     \partial_\nu \Bigg[h U^\mu U^\nu + \frac{h}{4n^2e^2}&J^\nu J^\mu+p\, \eta^{\mu\nu}\Bigg]=J_\nu F^{\mu\nu} \label{eq_momentum} \ , 
\end{align}
together with the generalized Ohm's law
\begin{align}
      \frac{1}{4ne}\partial_\nu \Bigg[\frac{h}{ne} \big(U^\mu J^\nu +&J^\mu U^\nu\big)\Bigg] \nonumber\\
    =U_\nu F^{\mu\nu}-\eta  \Bigg[J^\mu&+U_\alpha J^\alpha U^\mu \left(1+\Theta\right)\Bigg] \label{eq_ohm} \ . 
\end{align} 
In this set of equations, $h$ is the enthalpy of this relativistic plasma, $p$ is the pressure, $e$ the electron charge and $F^{\mu\nu}$ the electromagnetic tensor. Also, $\eta$ is the resistivity and $\Theta$ is the thermal energy exchange rate from negative to positive charged fluids \cite{koide2009}. A thermal function can be defined, depending only on the temperature $T$, such as $f=f\left(T\right)=h/\rho\geq 1$ for all temperature \cite{mahajan2003}.

In the pair plasma case, the Hall effect term does not appear in Ohm's law because of equal masses of the plasma species.
From Eqs. \eqref{eq_momentum} and \eqref{eq_ohm} it can be seen that thermal inertial effects  appear with terms proportional to the enthalpy. In the momentum equation they are associated to the inertia of current density, whereas in Ohm's law the thermal inertial effects correspond to the thermal electromotive force. 

Finally, along with the previous expressions, it is necessary to complement the plasma system with Maxwell's equations
\begin{equation}
    \partial_\nu F^{\mu\nu}=4\pi J^\mu\ , \hspace{0.8cm} \partial_\nu F^{*\mu\nu}=0\ , \label{eq_maxwell}
\end{equation}
where $F^{*\mu\nu}$ is the dual electromagnetic tensor.

The structure of the paper is as follows: In Sec. II, we present the theoretical model and governing equations for relativistic pair plasmas under asymmetric reconnection. In Sec. III, we explore the emergence of ultrarelativistic outflow velocities in highly asymmetric configurations. Finally, we summarize our results and discuss their implications for astrophysical plasma dynamics.


\section{Asymmetric reconnection model} \label{SPModel}

The above model is used to
describe an asymmetric magnetic reconnection process. For this,   physical vectorial quantities involved in the plasma dynamics are defined by projecting tensor them in the foliations of spacetime. For example, the electric and magnetic fields can be defined, respectively, as $E^\mu= n_\nu F^{\mu\nu}$ and $B^\mu= n_\nu F^{*\nu\mu}$, where $n^\mu=(1,0,0,0)$. Thus, the electric and magnetic field are defined by projections onto timelike hypersurfaces \cite{asenjocomisso2017}. Similarly, the four-velocity can be put in the form $U^\mu=\gamma  n^\mu+\gamma v^\mu$, such that Lorentz factor is defined as $\gamma=-n_\mu U^\mu/=(1-v_i v^i)^{-1/2}$, where we have considered the spacelike four velocity $v^\mu=(0,v^i)$, with
$v^i$ the spatial components of the vectorial plasma fluid  velocity \cite{asenjocomisso2017}.

\begin{figure}[h]\centering
\includegraphics[width=8.5cm]{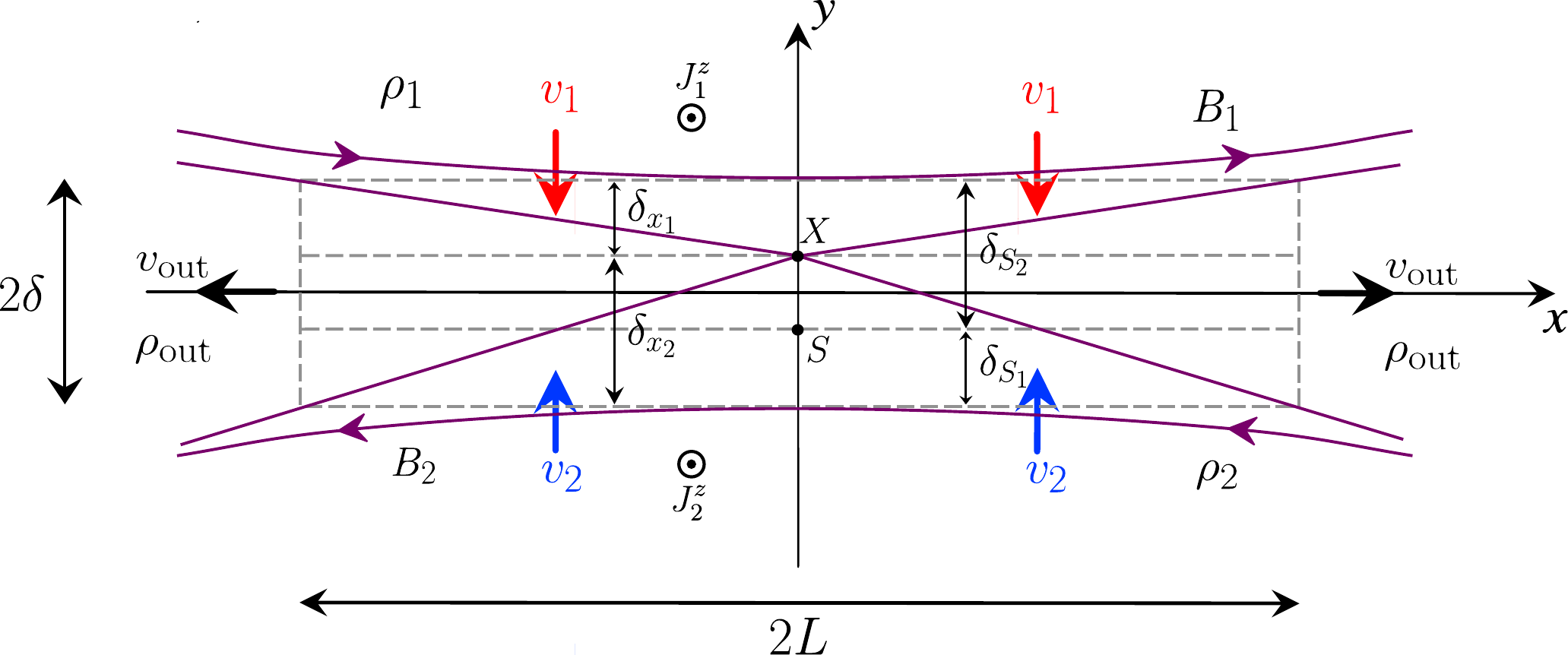}
\caption{Sweet-Parker reconnection configuration diagram for a relativistic asymmetric plasma. Subscripts "1" and "2" represent quantities above and below the dissipation region, respectively. Velocity flows are red and blue arrows, and the magnetic field is the purple line. The $S$ marks the stagnation point and the $X$ marks the X-line, both mobile.}\label{fig1}
\end{figure}

To estimate the magnetic reconnection rate with asymmetric inflow conditions in the Sweet-Parker configuration, we consider an asymmetric diffusion region of length $2L$ and width $2\delta$ (with $\delta\ll L$), as shown in Fig. \ref{fig1}.

When the magnetic reconnection process occurs at the steady state, the diffusion region has two different plasmas flows from above and below to this region, let us say, in the $y$-direction depicted in Fig. \ref{fig1}. Each of these plasma inflows can be described through Eqs.~\eqref{eq_cont} to \eqref{eq_ohm} in general, flowing with their own velocities, $v_1$ from above and $v_2$ from below, into the region of length $2L$ and width $2\delta$. Also, each plasma has its own density, $\rho_1$ and $\rho_2$, current density $J^{z}_1$ and $J^{z}_2$, magnetic field strength $B_1$ and $B_2$, and resistivity $\eta$, where from now on, sub indexes 1 and 2 stand for the above and below regions to the magnetic diffusion region, respectively. We also consider a reconnecting configuration such that $U_\mu J^\mu=0$ for both plasma inflows (see below).
Lastly, outside of the diffusion region, the plasma can be considered as ideal. 

After the magnetic reconnection takes place, the merged plasma is ejected from the diffusion region with symmetric outflow velocity $v_{\text{out}}$ and density $\rho_{\text{out}}$. In a steady state configuration, a relation between the inflow asymmetric velocities and the outflow velocities can be obtained through the continuity equation for the plasma. From Eq.~\eqref{eq_cont} we  estimate the rates for the conservation of flow as
\begin{equation} 
    2L\left(\rho_1\gamma_1 v_1+\rho_2\gamma_2 v_2\right)\sim 4\delta\rho_{\text{out}}\gamma_{\text{out}}v_{\text{out}}\ , \label{scale_cont}
\end{equation}
where $\gamma_1$, $\gamma_2$ and $\gamma_{\text{out}}$ are the Lorentz factors for plasma inflows 1 and 2, and  the plasma outflow, respectively. This is a relativistic generalization of the relation found in Ref.~\cite{CassakShay}.
Furthermore, the outflow density can be estimated with the help of the conservation of the magnetic flux in the reconnecting region.  The outflow density must be of the order of the the total effective density of the two inflow plasma flows \cite{CassakShay}. Thus,  
\begin{equation}
    \rho_{\text{out}}\sim \frac{\rho_1 B_2+\rho_2 B_1}{B_1+B_2}\, .\label{densitycassak}
\end{equation}

On the other hand, using Maxwell equations \eqref{eq_maxwell}, the momentum equation \eqref{eq_momentum} can be written as an expression for energy conservation. In the diffusion region, this equation acquires the form
\begin{eqnarray}
    &&2L\left[ \left(h_1+\frac{B_1^2}{4\pi}\right)\gamma_1^2 v_1+\left(h_2+\frac{B_2^2}{4\pi}\right)\gamma_2^2 v_2\right]\nonumber\\
    &&\qquad\qquad\qquad\quad\sim 4\delta \left(h_{\text{out}}+\frac{B_{\text{out}}^2}{4\pi}\right)\gamma_{\text{out}}^2 v_{\text{out}}\, ,
    \label{conservinicial}
\end{eqnarray}
where $h_{1,2,\text{out}}$ is the enthalpy in each region.

We can now divide Eqs.~\eqref{conservinicial} and \eqref{scale_cont} to obtain \cite{CassakShay}
\begin{eqnarray}
    &&\frac{\left(h_1+\sigma_1\rho_1\right)\gamma_1^2 v_1+\left(h_2+\sigma_2\rho_2\right)\gamma_2^2 v_2}{\rho_1\gamma_1v_1+\rho_2\gamma_2 v_2}\nonumber\\
    &&\qquad\qquad\qquad\quad\sim \left(h_{\text{out}}+\sigma_{\text{out}}\rho_{\text{out}}\right)\frac{\gamma_{\text{out}}}{\rho_{\text{out}}} \, ,
    \label{conservinicial2}
\end{eqnarray}
where we have defined the corresponding magnetization $\sigma_{1,2,\text{out}}\equiv B^2/4\pi \rho_{1,2,\text{out}}$, for each region.
Outside the diffusion region, the plasma follows an ideal Ohm's law \eqref{eq_ohm}, say $U_\nu F^{\mu\nu}=0$. For the current configuration, where the electric field is in a $z$-direction (out of the page), the ideal Ohm's law establishes that
\begin{equation}
   E^z\sim E_1^z\sim v_1 B_1\sim E_2^z\sim v_2 B_2\, .
   \label{ohmrelationideal}
\end{equation}
Using \eqref{ohmrelationideal} in Eq.~\eqref{conservinicial2}, we finally obtain
\begin{eqnarray}
  (f_{\text{out}}+\sigma_{\text{out}})  \gamma_{\text{out}}\sim \frac{(f_1+\sigma_1)\gamma_1^2+(f_2+\sigma_2)\xi\gamma_2^2}{\gamma_1+\xi\gamma_2}\, ,
   \label{relation1}
\end{eqnarray}
where $\xi\equiv \rho_2B_1/(\rho_1 B_2)$. The value of $v_{\text{out}}$ can be obtained through $\gamma_{\text{out}}$. In the non-relativistic cold limit,  $\gamma_1\sim 1$ , $\gamma_2\sim 1$, $f_1\sim 1$, $f_2\sim 1$, $f_{\text{out}}\sim 1$, $\gamma_{\text{out}}\sim 1+v_{\text{out}}^2/2$, and $\sigma_{\text{out}}$ is negligible, allowing us to obtain
the results of Ref.~\cite{CassakShay}. From \eqref{relation1}, we get $\gamma_{\text{out}}\sim 1+ B_1 B_2(B_1+B_2)/(4\pi(\rho_1 B_2+\rho_2 B_1))$, which allow us to obtain the non-relativistic value of $v_{\text{out}}$.

In a general fashion, the reconnection rate $E^z\sim v_1 B_1$ can be written using the continuity equation \eqref{scale_cont} as
\begin{eqnarray}
    E^z\sim \left(\frac{2\delta}{L}\right)\left(\frac{\rho_{\text{out}}\gamma_{\text{out}}v_{\text{out}}B_1 B_2}{\gamma_1 \rho_1B_2+\gamma_2\rho_2B_1} \right)\, .
    \label{reconrate1}
\end{eqnarray}

Both Eqs.~\eqref{relation1} and \eqref{reconrate1} are considered for a general relativistic
asymmetric configuration, since no diffusion mechanism has been made explicit yet. 
From the scheme of the interior configuration of the dissipation region shown in Fig. \ref{fig1} and Eq. \eqref{ohmrelationideal}, both $X$ and $S$ point are not fixed. This due to the asymmetry in velocity flows and magnetic fields from the regions above and below the diffusion area, making these points displaced from the center. In this way, according to \cite{CassakShay}, the length can be written as
\begin{eqnarray}
    2\delta=\delta_{X1}+\delta_{X2}=\delta_{S1}+\delta_{S2}\, ,
    \label{relationdeltas}
\end{eqnarray}
where $\delta_{X1}$ and $\delta_{X2}$ are the distances of each corresponding asymmetric region to the central X-line of reconnection. Similarly, $\delta_{S1}$ and $\delta_{S2}$ are the distances from the edges to the stagnation point.

On the other hand,  we now introduce a dissipation mechanism through Ohm's law \eqref{eq_ohm}. 
Notice that along the neutral line there is no contribution from the thermal energy exchange rate between the charged fluids. This due to that in our configuration $E^x\approx E^y\approx 0$ and $B^z\approx 0,$ while $J^0=0\approx J^x$ and $v^y\approx v^z\approx 0,$ implying that $U_\mu J^\mu \approx 0$ in this zone. 
Also, we consider that the thermal electromotive effects can be neglected in the reconnection layer, thus $\eta  J^\mu =\,  U_\nu F^{\mu\nu} $. 
In this way, the generalized Ohm's law \eqref{eq_ohm} yields $J^y=0$, and can be reduced in the diffusion region, for each inflow plasma, to 
\begin{align}
    \eta J^z_{1,2} &\sim   \gamma_\text{out} E^z_{1,2}\ , \label{eq-ohmg3}
\end{align}
where $J^{z}_{1,2}$ are the corresponding current densities  in the $z$-direction of the two asymmetric plasma inflows in the diffusion region, along the neutral line. Also, from Eq. \eqref{eq_maxwell}, Ampère's law, $\nabla\times B= 4\pi J$, is written for both inflow plasmas, allowing us to calculate the estimated contribution per region of the plasma current. In this way, we obtain
\begin{equation}
    J^{z}_{1,2}\sim \frac{B_{1,2}}{4\pi \delta_{X1,X2}}\, ,
    \label{relationJBdeltax}
\end{equation} 
written in terms of $\delta_X$ (a similar expression can be found in terms of $\delta_S$). 

Therefore, using  Eqs.~\eqref{ohmrelationideal}, \eqref{reconrate1}, \eqref{relationdeltas}, \eqref{eq-ohmg3}  and \eqref{relationJBdeltax}, we get relations for velocities $v_1$ and $v_2$ in the succinct form
\begin{eqnarray}
    v_{1}^2 \left(\rho_1 \gamma_1 \alpha + \rho_2 \gamma_2\right) &\sim& \frac{\alpha\left(1+\alpha\right)}{S} v_\text{out}\rho_{\text{out}}\, \label{defv1} ,\\
    v_{2}^2 \left(\rho_1 \gamma_1 \alpha + \rho_2 \gamma_2\right) &\sim& \frac{\left(1+\alpha\right)}{\alpha S} v_\text{out}\rho_{\text{out}}\, \label{defv2} \ . 
\end{eqnarray}
where $\alpha=B_2/B_1$ measures the asymmetry between magnetic field strengths, and $S=4\pi L/\eta$ is the relativistic Lundquist number. Eqs.~\eqref{defv1} and \eqref{defv2} are
coupled  for $v_1$ and $v_2$, depending on $\gamma_1$ and $\gamma_2$.

Finally, using  Eqs.~\eqref{ohmrelationideal}, \eqref{reconrate1},  \eqref{relationdeltas}, \eqref{defv1} and \eqref{defv2} we get the reconnection rate 
\begin{eqnarray}
    E^{z} &\sim& B_1 \sqrt{\frac{\alpha\left(1+\alpha\right)\rho_\text{out} v_\text{out}}{S \left(\rho_1 \gamma_1 \alpha + \rho_2 \gamma_2\right)}} \nonumber \ ,\\
    &\sim& B_2 \sqrt{\frac{\left(1+\alpha\right)\rho_\text{out} v_\text{out}}{\alpha S \left(\rho_1 \gamma_1 \alpha + \rho_2 \gamma_2\right)}} \label{reconrate2}\, .
\end{eqnarray}
 These results represent a relativistic generalization of the results of Cassak and Shay \cite{CassakShay}. The relativistic nature of the inflow plasma generates a new type of asymmetry that it does not have a counter-part in the non-relativist regime. This can be seen in the term $({1+\alpha})\rho_\text{out}/({ \rho_1 \gamma_1 \alpha + \rho_2 \gamma_2})$, which reduces to unity in the non-relativistic velocity limit ~\cite{CassakShay}. 

The above reconnection rate can be explicitly evaluated for different  cases by solving for the asymmetric velocities. However, a particular important case  
stands out that allows for ultrarelativistic plasma outflows, larger than the ones obtained for  symmetric reconnection cases.

\section{ultrarelativistic outflow velocities}\label{asy1sec}

Contrary to previous works \cite{mbarek2022,peery2024,figueiredo2024}, where almost-symmetric relativistic conditions are considered for magnetic reconnection process, here we focus in the opposite research direction. We explore the consequences of a highly asymmetric configuration. 

Let us consider the particular case of very asymmetric 
magnetic fields, $\alpha \gg 1$ ($B_2\gg B_1$). This also implies that $\sigma_2/\sigma_1\gg \rho_1/\rho_2$.
In here, we are interested in the specific magnetic reconnection process occurring for highly asymmetric inflow velocities, i.e., a fast inflow region with $\gamma_1\gg 1$ ($v_1 \lesssim   1)$, and a slow inflow region with $\gamma_2\sim 1$ ($v_2\lll 1$). This is consistent with Ohm's law \eqref{ohmrelationideal}.
Below, we show that in this scenario, the outflow plasma velocities become ultrarelativistic, larger than the inflow velocities.

We start by solving Eq. \eqref{defv1}, when $\gamma_1 \gg 1$, and under the condition
\begin{equation}
\rho_1 \gamma_1 \alpha \gg \rho_2 \gamma_2\sim \rho_2\, .   
\label{conditionUltra}
\end{equation}
Then, in such case, from Eq. \eqref{defv1} we obtain
\begin{eqnarray}
    v_{1}^2 \gamma_1 &\sim& \frac{\left(1+\alpha\right)}{S} \frac{\rho_{\text{out}}}{\rho_1} v_\text{out}\, , \label{ultrav1} 
\end{eqnarray}
As $\gamma_1\gg 1$, then $v_1^2\gamma_1=\gamma_1-1/\gamma_1\sim \gamma_1$. Thus, we finally get  a estimation for the  relativistic factor of the fast inflow region 
\begin{equation}
    \gamma_1 \sim \frac{\alpha\,  \rho_{\text{out}}}{S\rho_1} v_\text{out}, \, \label{gamma1}
\end{equation}
such that $v_1\sim 1-1/(2\gamma_1^2)$.
The $\gamma_1\gg 1$ requirement imposes the condition
\begin{eqnarray}
    \alpha\,  \rho_{\text{out}}v_\text{out}\gg {S\rho_1}\, . 
    \label{condUltra1}
\end{eqnarray}

We can now use this result in Eq.~\eqref{defv2}. By using condition \eqref{conditionUltra}, we get that $v_2^2\rho_1\gamma_1\alpha\sim \gamma_1/\alpha$, which   allow us to find that
\begin{eqnarray}
     v_{2} &\sim& \frac{1}{\alpha} \ . \label{valorv2}
\end{eqnarray}
Notice that as $\alpha\gg 1$, solution \eqref{valorv2} is consistent with $v_2\lll 1$ and $\gamma_2 \sim 1$. 
Solutions \eqref{gamma1} and \eqref{valorv2} cannot be reduced to the non-relativistic results of Ref.~\cite{CassakShay}.

With all the above results and conditions, we can estimate $\gamma_{\text{out}}$ from Eq. \eqref{relation1} using the found values for $\gamma_1$ and $\gamma_2$. We are particularly interested in finding an ultrarelativistic ourflow solution of the system, $\gamma_{\text{out}} \gg 1$ and $v_{\text{out}} \lesssim 1 $. In order to simplify the model, we examine the case where the outflow magnetization can be neglected, $\sigma_{\text{out}} \ll 1$, as well as we consider a cold plasma limit, $f_{1,2,\text{out}} \sim 1$. Then, as we can write $\xi = \rho_2/(\rho_1 \alpha)=\alpha\sigma_1/\sigma_2$, the outflow Lorentz factor becomes
\begin{eqnarray}
     \gamma_{\text{out}}  &\sim& \frac{(1+\sigma_1) \alpha^3\rho^2_{\text{out}}+(1+\sigma_2)S^2 \rho_1 \rho_2}{S  \rho_1\left[{\alpha^2\rho_{\text{out}}+S  \rho_2}\right]}\ .
   \label{gammaout}
\end{eqnarray}

On the other hand, using the  explicit value for $\gamma_1$ from Eq.~\eqref{gamma1} in condition \eqref{conditionUltra}, we get  
\begin{eqnarray}
\alpha^2 \rho_{\text{out}} \gg S \rho_2\, .
\label{condUltra2}
\end{eqnarray}
In addition, due to our assumptions, from Eq. \eqref{densitycassak} we obtain that $\rho_{\text{out}} =\alpha \rho_1 \left(1+\xi\right)/(1+\alpha) \sim \rho_1 \left(1+\xi\right)$.
Using this and condition \eqref{condUltra2}, the outflow Lorentz factor \eqref{gammaout} can be written as
\begin{eqnarray}
     \gamma_{\text{out}} &\sim& \frac{\alpha}{S}(1+\sigma_1)(1+\xi) + \frac{S\left( \xi+\alpha\sigma_1\right)}{\alpha(1+\xi)}\, , \label{gammaout2}
\end{eqnarray}
where we have used the fact that $\sigma_2=\alpha^2 \sigma_1\rho_2/\rho_1=\alpha \sigma_1/\xi$.

Besides this, the condition \eqref{condUltra1} implies that $\alpha/S\gg 1/(1+\xi)$, whereas condition \eqref{condUltra2} establishes that $\alpha/S\gg \xi/(1+\xi)$. Both of these conditions allow us to look for a regime where $\alpha/S\gg 1$ without infringing our assumptions. This is consistent with the large value of the  inflow Lorentz factor for the fast component. In fact, by  using the above results in \eqref{gamma1}, then we can re-write the  Lorentz factor  as
\begin{eqnarray}
    \gamma_1\sim \frac{\alpha}{S}(1+\xi)\, .
    \label{finalgamma1}
\end{eqnarray}

With help of \eqref{finalgamma1}, this outflow Lorentz factor
can be put as \begin{eqnarray}
\gamma_{\text{out}}\sim (1+\sigma_1)\gamma_1+\frac{\xi+\alpha\, \sigma_1}{\gamma_1}\, . 
\label{outflowLorentfinal1}
\end{eqnarray}
We consider the case of highly magnetized inflow $\sigma_1\gg 1$. In such case, the relation for the outflow Lorentz factor \eqref{outflowLorentfinal1} can be written as
\begin{eqnarray}
\frac{\gamma_{\text{out}}}{\sigma_1\gamma_1}\sim 1+\frac{S^2\sigma_2(1+\sigma_2)}{\alpha(\sigma_2+\alpha\sigma_1)^2}\, . 
\label{outflowLorentfinal2}
\end{eqnarray}
In the regime where $\sigma_2\ll \sigma_1$ (i.e. when $\rho_1/\rho_2\lll 1$) we find that $\gamma_{\text{out}}\sim {\sigma_1\gamma_1}$. On the contrary, when $\sigma_2\gg \sigma_1$ (allowing to have a large arbitrary range of values for $\rho_1/\rho_2$), then  $\gamma_{\text{out}}\sim {\sigma_1\gamma_1}(1+S^2/\alpha)$.

Therefore,  we find that $\gamma_{\text{out}} \gg \gamma_1\gg 1$, allowing an ultrarelativistic outflow plasma. The escaping plasma is accelerated by reconnection at larger relativistic velocities compared to the ones of the inflow plasma jets.
Also, we notice that $\gamma_{\text{out}}\gtrsim\sigma_1\gamma_1$, allowing  the asymmetric reconnection to have larger outflow velocities compared to those predicted by Lyutikov and Uzdensky for a relativistic symmetric process 
\cite{LyutikovUzdensky}. Finally, the reconnection rate \eqref{reconrate2} becomes simply $E^z\sim B_1$. 

The ultrarelativistic values for the resulting outflow plasma \eqref{outflowLorentfinal1} is the main result of this work. It establishes an effective mechanism for particle acceleration at high velocities  based in a highly asymmetric magnetic reconnection event. The 
ultrarelativistic outflow is achieved thanks to the energy available in the relativistic asymmetry of the reconnection process, and therefore, is not accessible in  symmetric reconnection. For instance, for $\alpha\sim 10^3$, asymmetric
values of magnetization $\sigma_1\sim 10^2$ and $\sigma_2\sim 0.1$, and a Lundquist number of about $S\sim 10^3$ \cite{Ripperda,Ripperda2,Nathanail}, then $\xi\sim 10^6$,  with $\rho_2/\rho_1\sim 10^9$. This implies a fast relativistic inflow $\gamma_1\sim 10^6$,  a slow inflow $v_2\sim 10^{-3}$ ($\gamma_2\sim 1$), and an ultrarelativistic outflow plasma with $\gamma_{\text{out}}\sim 10^8$. 

The above proposed asymmetric mechanism for ultrarelativistic outcomes from magnetic reconnection requires very specific (and somewhat difficult) conditions to occur. However, this result shows the large differences that a highly asymmetric reconnection process has compared with an almost-symmetrical one. It is our aim to show that the highly asymmetric limit is a regime that needs to be explored by the magnetic reconnection community.

\section*{Acknowledgments}
The authors of this work thank to
FONDECYT postdoc grant No. 3240441 (MAW), and to
FONDECYT grant No. 1230094 (FAA) for their support.



\end{document}